# Assessing Temporal Dispersion of Time-Varying Channel at 38 GHz for Different Reference Signals

Jarosław Wojtuń[1], Bartosz Czaja[1], Tomasz Grasza[1], Rafał Przesmycki[1], Jan M. Kelner[1], Cezary Ziółkowski[1], Radek Zavorka[2], Josef Vychodil[2], Tomas Mikulasek[2], Jiri Blumenstein[2], Ales Prokes[2], and Aniruddha Chandra[3]

[1] Institute of Communications Systems, Faculty of Electronics, Military University of Technology, Warsaw, Poland, *jaroslaw.wojtun@wat.edu.pl*

[2] Department of Radio Electronics, Brno University of Technology, Brno, Czech Republic

[3] ECE Department, National Institute of Technology, Durgapur, India[1]

***Abstract*—This paper investigates the influence of reference waveform selection on temporal dispersion estimation in a time-varying radio channel at 38 GHz. Three reference waveforms—numerically generated in MATLAB, measured through a conducted connection, and recorded in an anechoic chamber—are compared using the same wideband channel measurements conducted in a dynamic road environment. Their effects on the estimated power delay profiles (PDPs), mean delay, and root mean square (RMS) delay spread are evaluated. Despite noticeable differences in the time-domain and spectral characteristics of the reference waveforms, the resulting PDPs exhibit similar multipath structures, while the estimated temporal dispersion parameters show relatively small differences. The results indicate limited sensitivity to reference waveform selection under the investigated measurement conditions.**

***Index Terms*—propagation, millimeter-wave, temporal dispersion, time-varying channel, measurements, power delay profile, RMS delay spread, reference signal.**

## I. Introduction

Temporal dispersion is one of the fundamental characteristics of wideband radio propagation channels, describing the spreading of transmitted signals in the delay domain due to multipath propagation. Its characterization is particularly important in dynamic environments, where the motion of transmitters, receivers, or surrounding objects changes the multipath structure. The power delay profile (PDP) represents the received signal power distribution over propagation delay and is commonly derived from the estimated channel impulse response (CIR). Based on the PDP, the mean delay and root mean square (RMS) delay spread can be determined to characterize the temporal dispersion of the channel [1], [2]. Among these parameters, RMS delay spread is particularly important for assessing frequency selectivity, intersymbol interference, and the performance of wideband communication systems.

In wideband channel sounding measurements, the CIR can be estimated by cross-correlating the received signal with a known sounding waveform. For discrete-time signals, the estimated CIR is expressed as

$$h(\tau_k) = \sum_{n=0}^{N-1} r(t_n) s^*(t_n - \tau_k), \qquad (1)$$

where $r(t_n)$ denotes the received signal, $s(t_n - \tau_k)$ is the time-shifted sounding waveform, and $(\cdot)^*$ denotes complex conjugation. The sampling instants and discrete delays are defined as $t_n = n/f_s$, and $\tau_k = k/f_s$, respectively, where $f_s$ is the sampling frequency and $N$ is the number of received samples.

The resulting PDP is given by $P(\tau_k) = |h(\tau_k)|^2$. For a measured discrete PDP, the RMS delay spread, $\sigma_\tau$, is defined as [1]

$$\sigma_\tau = \sqrt{\frac{1}{\bar{P}} \sum_{k=1}^{K} P_k (\tau_k - \bar{\tau})^2}, \qquad (2)$$

where $\bar{P} = \sum_{k=1}^{K} P(\tau_k)$ denotes the total power of the retained PDP samples and $\bar{\tau} = \frac{1}{\bar{P}} \sum_{k=1}^{K} P(\tau_k)\tau_k$ represents the mean delay, while $P(\tau_k)$, and $\tau_k$ denote the power and propagation delay of the $k$th PDP sample, respectively.

Consequently, the estimated RMS delay spread depends not only on the physical multipath characteristics but also on the shape and power distribution of the estimated PDP.

The estimated CIR depends on the representation of the sounding waveform used in the correlation process, even when the same transmitted signal is considered. The waveform $s(t)$ may be represented by ideal complex in-phase and quadrature (IQ) samples generated in a numerical environment, such as MATLAB. Alternatively, it may be obtained from an actual measurement through a conducted connection with an attenuator, thereby accounting for the combined response of the hardware components included in the measurement path, such as amplifiers, cables, and adapters. Another possibility is to record the waveform in a controlled radiated environment, such as an anechoic chamber, which additionally includes the effects of the transmitting and receiving antennas. These approaches provide different representations of the sounding waveform and may therefore lead to differences in the estimated CIR, even when the same received signal is processed.

The influence of channel sounding equipment and calibration procedures on the estimated channel characteristics has been investigated in several studies. Gentile *et al.* [3] demonstrated that differences in channel

This research was funded in part by the National Science Center (NCN), Poland, grant no. 2021/43/I/ST7/03294 (MubaMilWave).

sounder architectures, hardware characteristics, and processing methods can affect the estimation of multipath components. Li *et al.* [4] examined amplitude and phase calibration in wideband channel sounding, showing that hardware imperfections may introduce artificial multipath components into the CIR. Papazian *et al.* [5] addressed impulse-response distortions caused by internal reflections and radio-frequency components using back-to-back calibration. Jun *et al.* [6] investigated over-the-air calibration of a 28 GHz phased-array channel sounder, demonstrating the importance of antenna-dependent calibration for reducing unwanted components in the measured impulse response. Quimby *et al.* [7] further showed that hardware imperfections, calibration methods, and signal processing can affect the resulting PDP and RMS delay spread estimates.

Although these studies demonstrate the importance of hardware characterization and calibration, the influence of the waveform selected for cross-correlation on the estimated temporal dispersion parameters requires further experimental assessment. In particular, differences between numerically generated and experimentally recorded sounding waveforms may modify the correlation response and consequently the estimated PDP, even when the propagation measurements remain unchanged. This paper investigates the influence of reference waveform selection on RMS delay spread estimation using wideband channel sounding measurements [8]conducted at 38 GHz in a dynamic road environment. The considered quasi-monostatic scenario includes stationary scatterers and moving vehicles, producing pronounced variations in the multipath delay structure. The analysis compares CIRs obtained from the same recorded data using different sounding waveforms, with particular attention to the resulting PDPs and RMS delay spread estimates. The main contribution is an experimental assessment of the sensitivity of temporal dispersion estimates to the choice of waveform used for CIR estimation in a vehicle-induced, time-varying millimeter wave (mmWave) propagation environment.

The remainder of this paper is organized as follows. Section II describes the measurement setup and the investigated propagation scenario. Section III presents the sounding waveforms used for CIR estimation and their characteristics. Section IV compares the resulting PDPs and RMS delay spread estimates. Finally, Section V summarizes the main findings and conclusions.

## II. Measurement Campaign

### A. Measurement Setup

The measurement campaign employed a wideband millimeter-wave (mmWave) channel sounder comprising spatially separated transmitting (Tx) and receiving (Rx) units, as illustrated in Fig. 1. The system was configured to investigate the temporal evolution of radio propagation characteristics in the presence of moving vehicles. The Tx and Rx subsystems were based on a Keysight M9384B VXG vector signal generator and a Keysight N9032B PXA signal analyzer, respectively. Both instruments support operation at frequencies up to 44 GHz and instantaneous bandwidths of up to 1 GHz. Time and frequency synchronization was provided by a Stanford Research Systems FS740 GPS-disciplined reference.

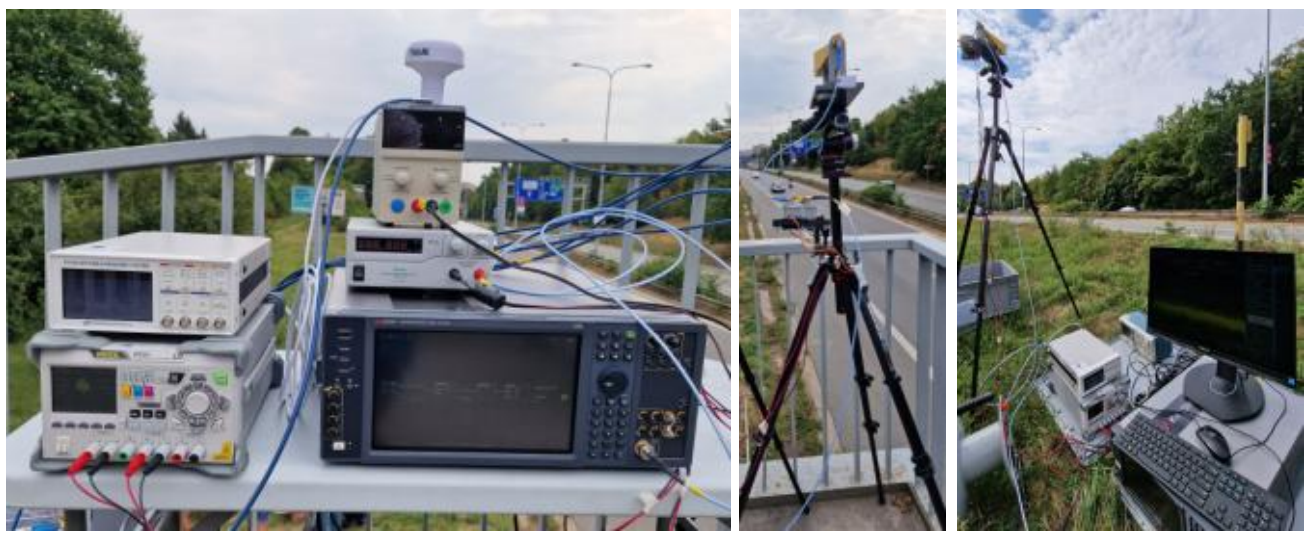

(a) (b) (c)

Fig. 1. Measurement testbed: (a) TX side, (b) TX antenna, (c) RX side.

The Tx chain incorporated an Eravant SBP-2034434526-2F2F-S1 Ka-band power amplifier (PA), providing a nominal gain of 45 dB over the 20–44 GHz frequency range. The signal generator was connected to the PA through a 5 m low-loss coaxial cable, while a 0.5 m cable connected the amplifier to the Tx antenna. At the Rx end, the received signal was amplified using an Eravant SBL-2434434025-2F2F-E3 Ka-band low-noise amplifier (LNA), with a nominal gain of 40 dB over the 24.25–43.5 GHz frequency range, and subsequently recorded by the PXA signal analyzer. The Rx antenna–LNA and LNA–analyzer connections employed 0.5 m and 5 m low-loss coaxial cables, respectively. The acquired signals were stored on a computer for subsequent processing.

Both subsystems employed identical Eravant SAR-2507-28-S2 rectangular horn antennas operating over the 26.5–40 GHz frequency band. The antennas provide a nominal gain of 25 dBi and half-power beamwidths (HPBWs) of approximately 7° in elevation and 9° in azimuth, enabling directional illumination and reception within the investigated road section [1], [9].

Channel sounding was performed using a frequency-modulated continuous-wave (FMCW) chirp with a bandwidth of 1 GHz and a duration of 8 µs. The sampling rates of the Tx and Rx instruments were set to 1.2 GSa/s. On the Rx side, a Tektronix AFG3021 external function generator triggered the acquisition of 16 µs recording windows every 25 ms, corresponding to a channel observation rate of 40 Hz. Each window contained at least one complete FMCW chirp suitable for subsequent channel estimation. A 30 s measurement session yielded 1,200 acquisition windows [8], [10]. All measurements analyzed in this paper were conducted at a carrier frequency of 38 GHz.

### B. Measurement Scenario

The measurements were conducted on August 26, 2026, along Hradecká Street (road 640/E461), near the Brno University of Technology (BUT) campus in Brno, Czech Republic. The investigated location provided a relatively open propagation environment with regular vehicular traffic and a limited number of stationary scattering objects. Fig. 2 depicts an aerial view of the measurement location and the deployment of the Tx and Rx units [8], [10].

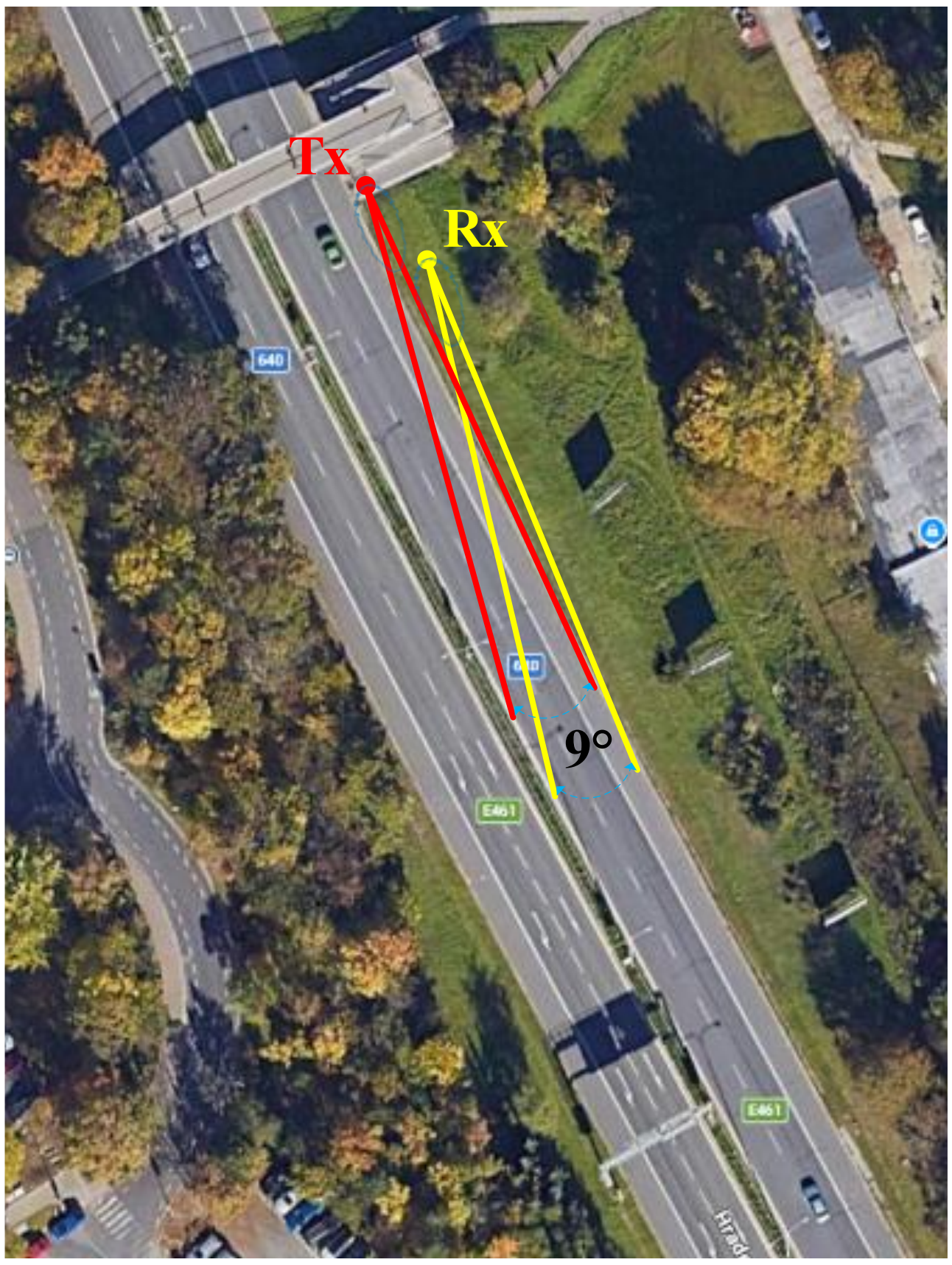

Fig. 2. Aerial view of measurement scenario along Hradecká Street near BUT campus, showing Tx and Rx locations and monitored traffic direction (background map: Google Maps).

The transmitter was installed on a pedestrian footbridge crossing Hradecká Street between Technická and Purkyňova Streets, with the Tx antenna positioned approximately 5 m above road level. The receiver was deployed near the roadside, slightly displaced from the transmitter along the monitored road section, with its antenna mounted at a height of approximately 1.5 m.

Both horn antennas were oriented approximately along the same direction, with their main beams aligned with the nearest traffic lane and facing approaching vehicles. Since the Tx–Rx separation was relatively small compared with the distances to the observed vehicles, the measurement geometry can be regarded as quasi-monostatic. The direct propagation path between the antennas remained outside their main beams, and the investigated vehicle-related propagation components were therefore observed under non-line-of-sight (NLOS) conditions.

The surrounding environment contained several stationary scatterers, including street lighting poles, traffic signs, overhead road information boards, and metallic safety barriers separating opposite traffic directions. These objects contributed relatively persistent multipath components, whereas vehicles passing through the illuminated road section introduced time-varying reflections and scattering. Depending on the instantaneous traffic conditions, different numbers and types of vehicles contributed to the received signal, producing variations in the CIR, PDP, and associated temporal dispersion characteristics [8], [10].

The resulting measurements provide a basis for investigating how different reference signals used in correlation-based channel estimation influence the reconstructed CIR, PDP, and RMS delay spread in a time-varying mmWave propagation environment.

## III. Sounding Waveform Representations

Three reference waveforms were considered for correlation-based CIR estimation: a numerically generated FMCW chirp, a waveform measured through a conducted RF connection with an attenuator, and a waveform recorded in an anechoic chamber. All three representations correspond to the same nominal sounding signal, with a duration of 8 µs and a bandwidth of 1 GHz, but differ in the extent to which they incorporate the response of the measurement system.

The first reference was generated numerically in MATLAB using the nominal FMCW signal parameters. It represents an idealized sounding waveform without distortions introduced by the RF hardware. The second reference was obtained from a conducted measurement through an attenuated connection, thereby incorporating the combined response of the measurement components included in the conducted signal path. The third reference was recorded in an anechoic chamber and additionally includes the effects of the radiating and receiving antennas and the corresponding measurement geometry.

Fig. 3 compares the three reference waveforms in terms of their time-domain magnitudes, magnitude spectra, and spectrograms. The MATLAB-generated waveform exhibits an approximately constant amplitude envelope and a relatively uniform spectrum over the occupied bandwidth. Its spectrogram reveals the expected linear frequency sweep of the FMCW chirp.

In contrast, the experimentally recorded references exhibit amplitude variations and nonuniform spectral characteristics. The conducted reference shows pronounced variations in its amplitude envelope and additional features in the spectrogram, whereas the anechoic-chamber reference exhibits amplitude fluctuations and a more distributed low-level spectral background. Nevertheless, the principal FMCW chirp trajectory remains visible in all three representations.

These differences indicate that the measured reference waveforms contain additional characteristics of the measurement system that are absent from the ideal MATLAB-generated signal. Since the correlation-based CIR estimation depends on the selected reference waveform, differences in its amplitude and spectral characteristics may affect the resulting CIR, PDP, and temporal dispersion parameters. To evaluate this influence, the same received signal recordings were processed using each of the three reference waveforms, ensuring that any differences in the estimated channel parameters originate from the reference waveform selection rather than changes in the measured propagation scenario.

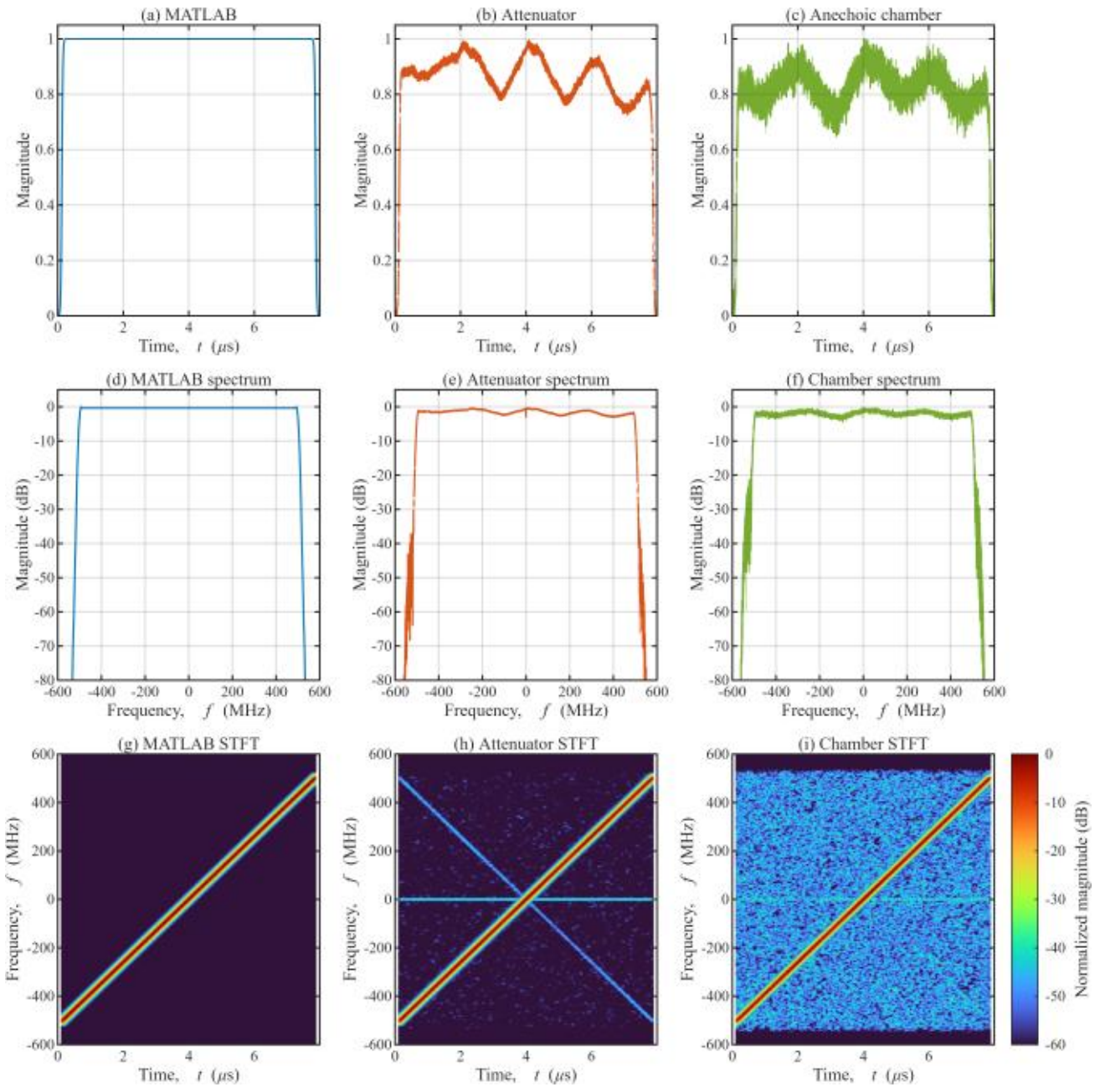

Fig. 3. Comparison of three FMCW waveform representations: MATLAB-generated, conducted through an attenuator, and recorded in an anechoic chamber. (a)–(c) Time-domain magnitudes, (d)–(f) magnitude spectra, and (g)–(i) spectrograms.

## IV. Results

The influence of reference waveform selection on temporal dispersion characterization was evaluated using the same 30 s measurement record acquired at 38 GHz. The received signals were cross-correlated with each of the three reference waveforms described in Section III. The resulting CIRs were used to calculate the corresponding PDPs, mean delays, and RMS delay spreads.

Fig. 4 compares the PDPs obtained using the MATLAB-generated, conducted, and anechoic-chamber reference waveforms. All three profiles exhibit similar multipath structures, including persistent horizontal traces associated with stationary scatterers and inclined trajectories corresponding to moving vehicles. The downward-sloping trajectories primarily represent approaching vehicles, whereas upward-sloping trajectories are associated with receding vehicles. Their consistent positions across the three PDPs indicate that the dominant propagation paths are identified similarly regardless of the selected reference waveform.

Nevertheless, differences in the relative intensities and background levels of individual multipath components can be observed. These variations are consistent with the differences in the amplitude and spectral characteristics of the reference waveforms shown in Fig. 3. Consequently, although the overall delay structure remains similar, the power distribution within the estimated PDP depends to some extent on the reference waveform used for correlation processing.

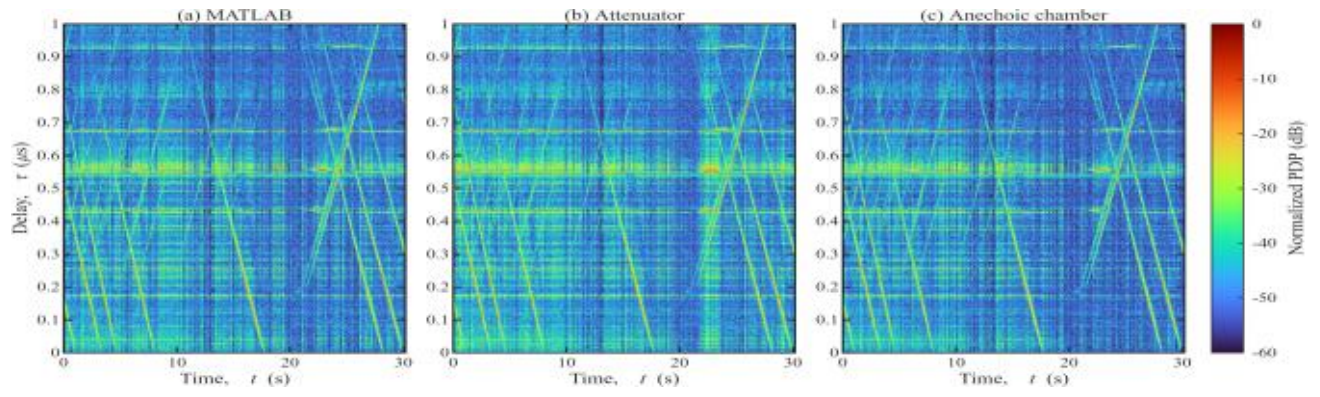

Fig. 4. Comparison of three PDPs generated with different reference signals: MATLAB-generated (a), conducted through an attenuator (b), and recorded in an anechoic chamber (c).

To assess the influence of these differences on temporal dispersion estimates, the mean delay and RMS delay spread were calculated for each observation using a relative PDP power threshold of −30 dB. Fig. 5 illustrates their temporal evolution over the 30 s measurement interval. Both parameters exhibit pronounced time variations associated with changes in the instantaneous multipath structure as vehicles enter and leave the monitored road section.

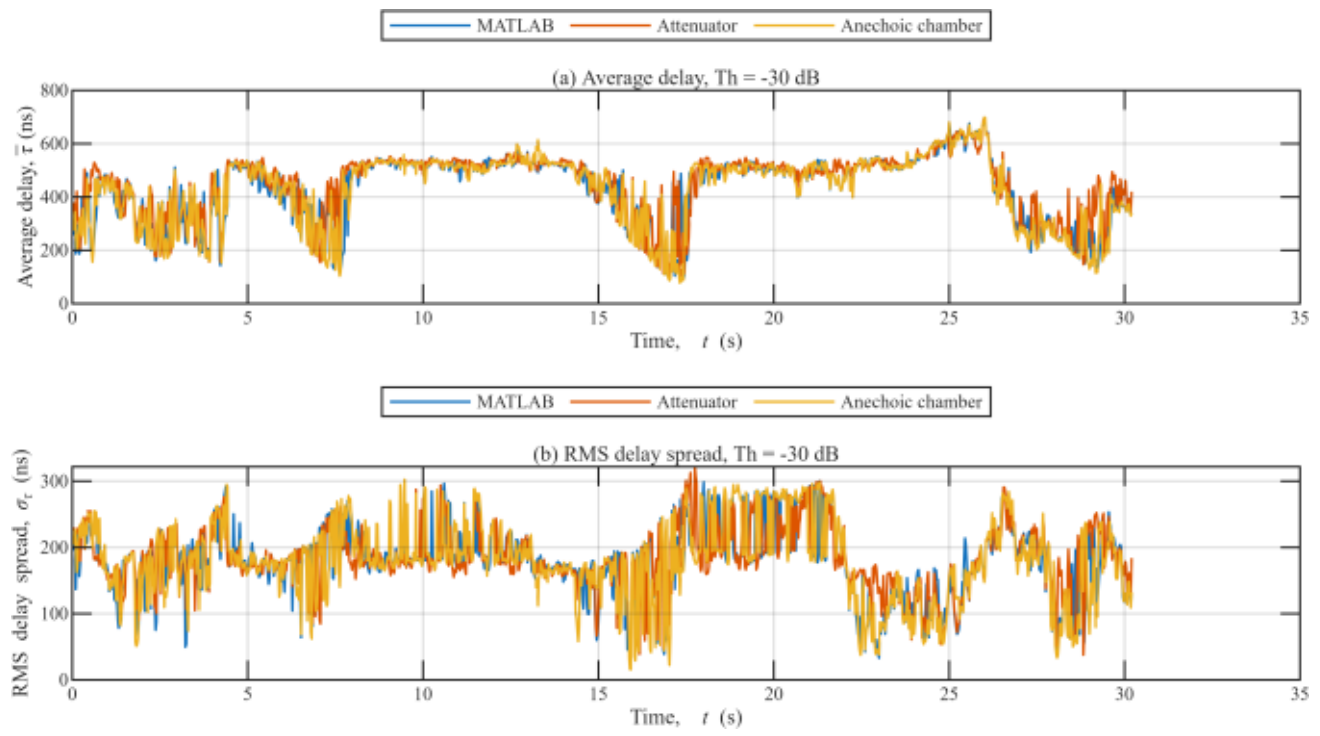

Fig. 5. Comparison of average delay (a) and RMS delay spread (b) parameter values with regard to different reference signals.

The three reference waveforms produce closely aligned temporal trajectories for both mean delay and RMS delay spread. The main increases and decreases occur at approximately the same observation times, indicating that reference selection does not substantially alter the temporal evolution of the estimated propagation characteristics. However, small differences between the curves are visible, particularly during intervals of rapid variation, when the relative powers of individual multipath components change.

Fig. 6 depicts box plots summarizing the distributions of mean delay and RMS delay spread for the three reference waveforms. The median values and interquartile ranges are broadly comparable, indicating that the overall statistical characteristics of the estimated parameters remain similar. Differences are more noticeable in the lower distribution tails and among the outliers, suggesting that reference waveform selection has a greater influence on some individual channel observations than on the central tendency of the estimated parameters.

The empirical cumulative distribution functions (CDFs) shown in Fig. 7 provide a complementary statistical comparison. The mean delay distributions exhibit small differences, particularly in their lower-value regions, while converging toward similar values at higher percentiles. The RMS delay spread distributions also largely overlap, with modest differences in their lower and intermediate ranges. These observations are consistent with the temporal trajectories and box plots, confirming that the influence of reference waveform selection is relatively limited for the investigated measurement scenario.

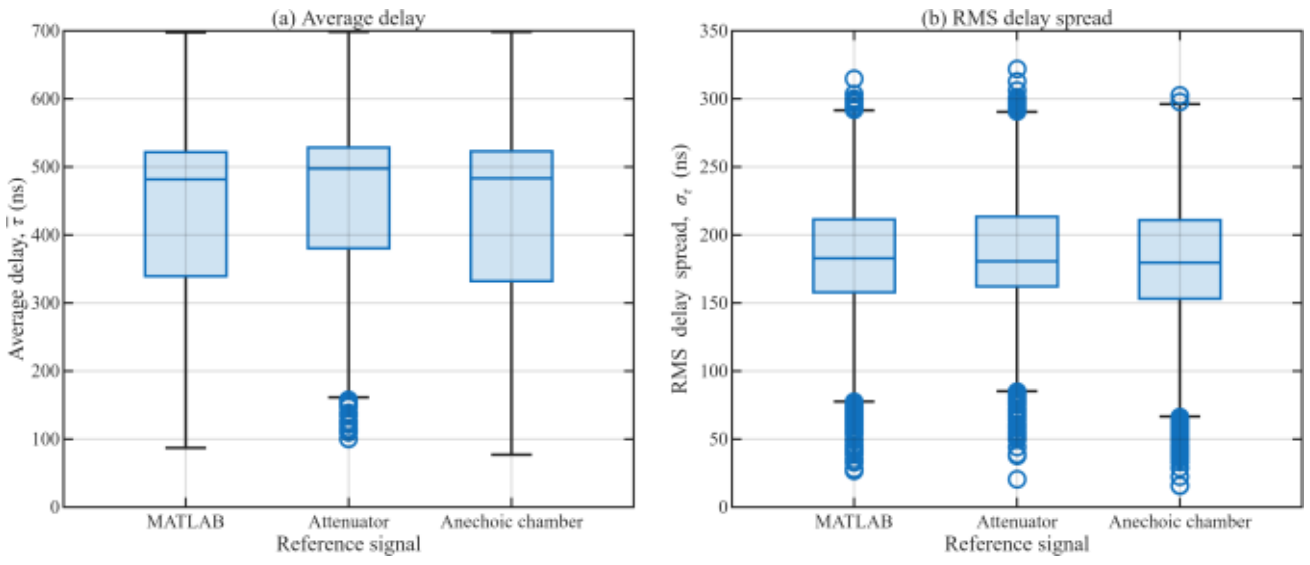

Fig. 6. Box plots of Average delay (a) and RMS delay spread (b) parameter values from CIRs computed with 3 different reference signals.

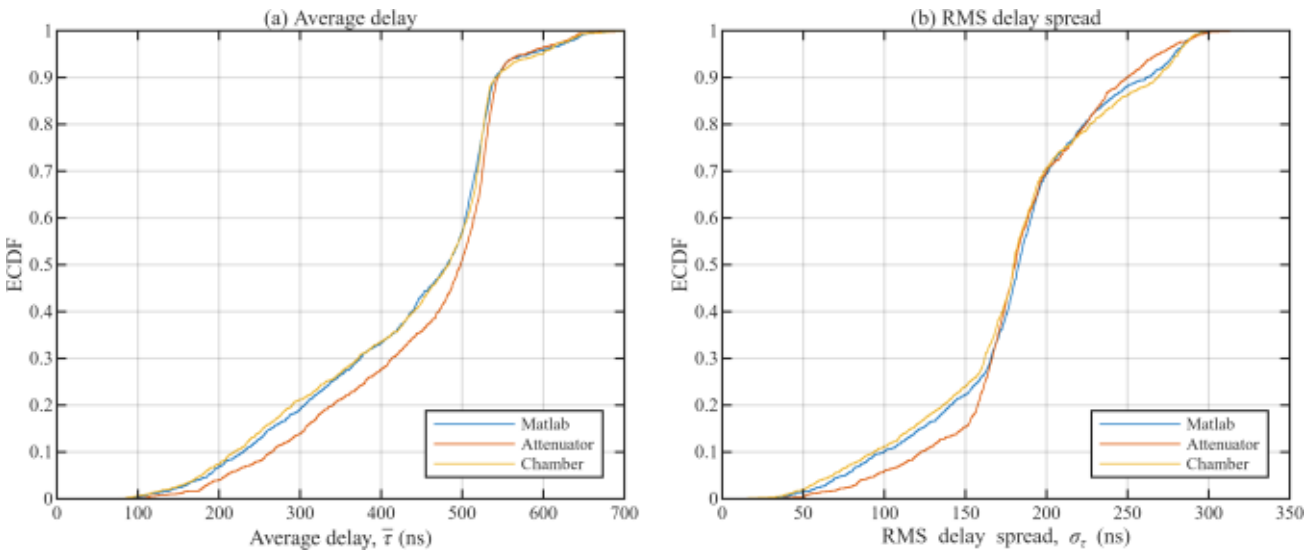

Fig. 7. Empirical Cumulative Distribution Functions of Average delay (a) and RMS delay spread (b) parameter values from CIRs computed with 3 different reference signals.

Overall, the results demonstrate that the choice of reference waveform affects the estimated PDP through differences in the correlation response, but its influence on the resulting mean delay and RMS delay spread statistics is comparatively small. The measured reference waveforms incorporate additional characteristics of the measurement system, whereas the numerically generated reference represents an idealized sounding signal. Nevertheless, all three approaches yield broadly consistent temporal dispersion estimates for the investigated time-varying channel at 38 GHz. The observed differences should therefore be interpreted as sensitivity to the reference waveform selection rather than evidence of superior estimation accuracy for any particular approach.

## V. Conclusion

This paper investigated the influence of reference waveform selection on temporal dispersion estimation in a time-varying radio channel at 38 GHz. Three reference waveforms—numerically generated, measured through a conducted connection, and recorded in an anechoic chamber—were compared using the same wideband channel measurements. Despite noticeable differences in their time-domain and spectral characteristics, all three references produced similar PDPs and broadly consistent mean delay and RMS delay spread estimates. The observed differences were more pronounced for individual channel observations than for the overall statistical distributions. These findings indicate that temporal dispersion estimates exhibit relatively limited sensitivity to reference waveform selection under the investigated measurement conditions.

## Acknowledgment

This work was co-funded by the National Science Centre, Poland, under the OPUS call in the Weave programme under research project no. 2021/43/I/ST7/03294 acronym 'MubaMilWave'; by the Czech Science Foundation under project no. 23-04304L; by the Military University of Technology under project no. UGB 531-000128-W300-22/2026 "Analysis of the possibility of using modern technologies in communication systems and electronic warfare"; and by the MeitY, Government of India, through the Chips-to-Startup (C2S) program under project no. EE-9/2/2021-R&D-E.